\documentclass[
superscriptaddress,
nofootinbib,
 amsmath,amssymb,
 aps,
pra,
twocolumn]{revtex4-2}

\usepackage[utf8]{inputenc}
\usepackage{graphicx}
\usepackage{hyperref}
\hypersetup{colorlinks = true, linkcolor = [rgb]{0.19411,0.51882,0.667058}, urlcolor = [rgb]{0.125490,0.29542,0.1647058}, citecolor = [rgb]{0.75882,0.37411,0.14117}, breaklinks=true}
\usepackage{amsmath}
\usepackage{amsfonts}
\usepackage{amssymb}
\usepackage{dsfont}
\usepackage{braket}
\usepackage{comment}
\usepackage[format=plain]{caption}
\usepackage{subcaption}

\begin{document}

\title{Loss-aware pulse sequence optimization for generating photonic Fock states}

\author{Benjamin Stodd}
\affiliation{Institute of Condensed Matter Theory and Optics, Friedrich Schiller University Jena, Max-Wien-Platz 1, 07743 Jena, Germany}
\author{Priyanshu Tiwari}
\affiliation{Heinz Nixdorf Institute, Paderborn University
Fürstenallee 11, 33102 Paderborn, Germany
}
\affiliation{Department of Electrical Engineering and Information Technology
Paderborn University, Warburger Str. 100, 33098 Paderborn, Germany}
\affiliation{
Institute of Applied Physics, Abbe Center of Photonics,
Friedrich Schiller University Jena, Albert-Einstein-Straße 15, 07745 Jena, Germany}
\author{René Sondenheimer}
\affiliation{Institute of Condensed Matter Theory and Optics, Friedrich Schiller University Jena, Max-Wien-Platz 1, 07743 Jena, Germany}
\affiliation{Fraunhofer Institute for Applied Optics and Precision 
Engineering IOF, Albert-Einstein-Str. 7, 07745 Jena, Germany}
\author{Sina Saravi}
\affiliation{Heinz Nixdorf Institute, Paderborn University
Fürstenallee 11, 33102 Paderborn, Germany
}
\affiliation{Department of Electrical Engineering and Information Technology
Paderborn University, Warburger Str. 100, 33098 Paderborn, Germany}
\affiliation{Institute for Photonic Quantum Systems (PhoQS)
Paderborn University, Warburger Str. 100, 33098 Paderborn, Germany}
\affiliation{
Institute of Applied Physics, Abbe Center of Photonics,
Friedrich Schiller University Jena, Albert-Einstein-Straße 15, 07745 Jena, Germany}
\author{Martin G\"{a}rttner}\email{martin.gaerttner@uni-jena.de}
\affiliation{Institute of Condensed Matter Theory and Optics, Friedrich Schiller University Jena, Max-Wien-Platz 1, 07743 Jena, Germany}

\begin{abstract}
We investigate the preparation of frequency-tunable photonic Fock states in a hybrid cavity system consisting of a nonlinear medium and a two-level system. Employing a gradient-based optimization approach, we construct multipulse driving protocols that control the system dynamics through pulse amplitudes, phases, and inter-pulse delays. Assuming unitary dynamics, the optimized sequences enable near-deterministic preparation of low-photon-number Fock states. We extend the optimization framework to open-system dynamics by modeling atomic decay and photon loss within the Lindblad master equation. This allows us to identify pulse sequences that exhibit enhanced robustness against dissipation compared to those optimized under idealized assumptions. Furthermore, we find that optimal pulse sequences obey strict constraints on relative phases, which are limited to values of 0 or 
$\pi$. These phase restrictions are supported by an analytical study that investigates a simple two-pulse sequence treating the
second pulse perturbatively.
\end{abstract}

\date{\today}

\maketitle

\section{Introduction}

\begin{figure}
    \centering
    \includegraphics[width=0.9\linewidth]{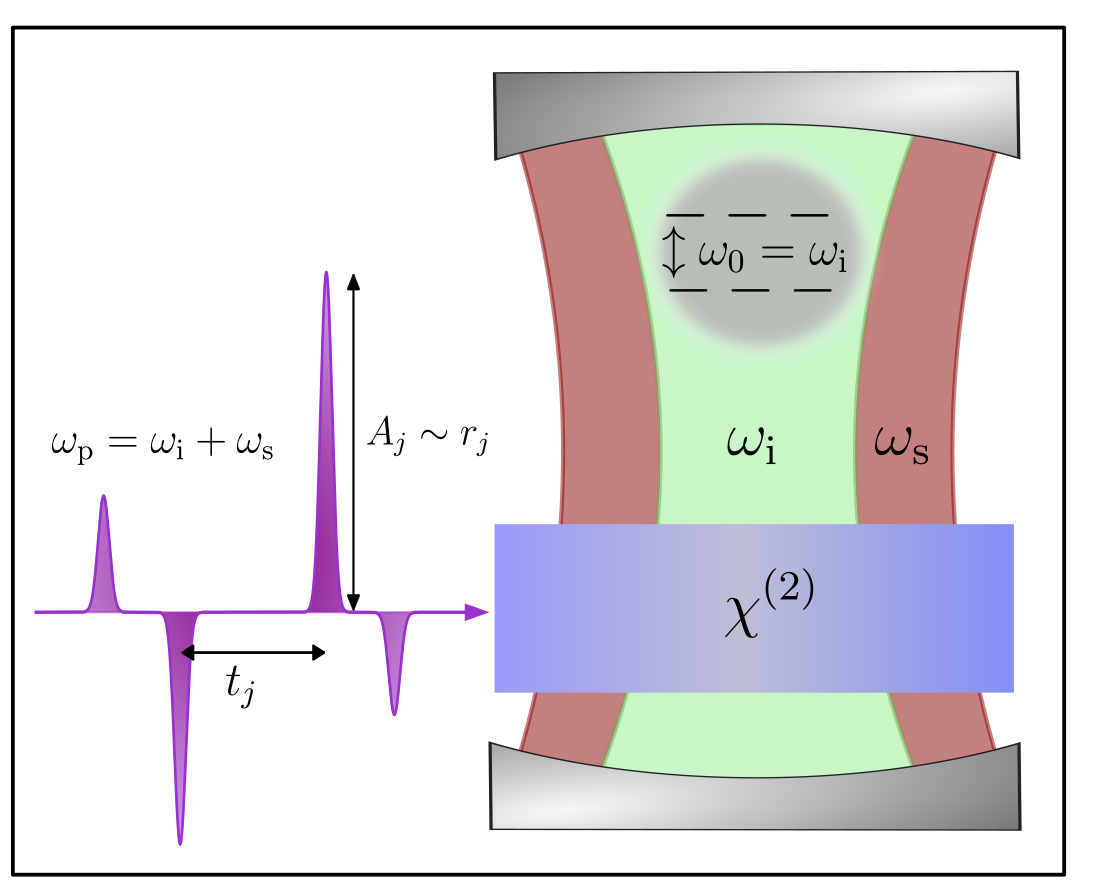}
    \caption{Illustration of the hybrid system. A nonlinear crystal with second-order susceptibility
$\chi^{(2)}$ is driven by an optical field of central frequency $\omega_\mathrm{p} = \omega_\mathrm{i} +\omega_\mathrm{s}$. One of the cavity modes is
resonant with the transition frequency of the 2LS inside the cavity, $\omega_\mathrm{i} = \omega_\mathrm{0}$. The amplitude of the $j$th pulse is proportional to the parametric gain $r_j$.}
\label{fig:hybrid_system_image}
\end{figure}

The preparation of non-classical states of
light is a central challenge in quantum information theory and essential for the realization of optical quantum technologies \cite{Flamini2018,Weedbrook2012, Brod2019, PerarnauLLobet2020}. In particular, the preparation of photonic Fock states with photon numbers larger than one poses major experimental
difficulties.
Usually, it is endeavored in two different ways. One approach relies on atomic or solid-state emitters with atom-like properties \cite{Uria2020, Brown2003, GonzalesTudela2015, Cosacchi2020}, where successful implementations have mainly been realized by employing temporal demultiplexing of single photons emitted from a single quantum dot \cite{Lenzini2016, Hansen2023, Wang2019}. Another approach to Fock state preparation uses nonlinear parametric sources, enabling the heralding of specific number states via conditional measurements on squeezed light, which can be generated using
nonlinear parametric down-conversion  \cite{Cooper2013, Tiedau2019, Waks2006}. Both methods exhibit advantages and disadvantages. While atomic systems provide deterministic generation but limited possibilities for engineering spectral and modal properties of the obtained state of light \cite{Lee2020}, nonlinear sources offer high tunability \cite{Wang2021} but only probabilistic generation \cite{Tiedau2019}. As the two main methods for Fock state generation offer complementary strengths, this naturally motivates the combination of the two kinds of systems in a hybridized fashion.\\
\indent Recently, hybrid quantum optical systems combining a two-level system (2LS) with a pump-pulse driven nonlinear crystal in an optical cavity have been proposed for high-fidelity Fock state preparation \cite{Krstic2024}. An illustration of this setup is given in Fig.~\ref{fig:hybrid_system_image}. 
In the hybrid system, the cavity supports two non-degenerate modes, denoted as the signal and idler modes, with frequencies $\omega_\mathrm{s}$ and $\omega_\mathrm{i}$. While the idler mode is resonant with the atomic transition frequency $\omega_0$ of the 2LS, the signal mode remains far detuned. The dynamics are governed by two interaction mechanisms. The pump pulses driving the nonlinear crystal induce a two-mode squeezing interaction, generating pairs of photons in the signal and idler modes. Further, the resonant coupling between the idler mode and the 2LS gives rise to Rabi oscillations. Carefully engineered pump-pulse amplitudes, phases and inter-pulse delays can drive the state of the composite system into a state where the signal mode has a high fidelity with respect to a target Fock state. The frequency of the signal mode can be fully controlled by tuning the frequency spectrum of the pump pulses and the
resonances of the nonlinear cavity, while the 2LS can remain
unchanged \cite{Krstic2024}. Thus, this hybrid approach enables
the preparation of Fock states at any desired frequency, while also offering near on-demand preparation capability.\\ \indent However, in the previous work, the reachable preparation
fidelities decreased rather quickly with the photon number. While it was shown that simple pulse sequences of up to three pulses can achieve near-deterministic preparation of single-photon states, optimal Fock state preparation for $N \geq 2$ is limited by the restricted control parameter space accessible in few-pulse protocols. In addition to this, effects of dissipation have been discussed on the preparation-process corresponding to pulse sequences that were optimized under ideal conditions. But a loss-aware optimization framework of the hybrid system has not been introduced. For multi-pulse sequences consisting of four or more pulses, the rapidly increasing control-parameter space requires an advanced optimization framework. At the same time, incorporating loss mechanisms substantially increases the computational cost of the system modeling, making an efficient numerical implementation essential.\\
\indent While many techniques for optimizing quantum state preparation in dynamical systems have been developed based on gradient-based methods \cite{KHANEJA2005296, Brif_2010, Ansel_2024}, in this work we introduce a model-specific gradient-descent scheme that enables efficient exploration of the high-dimensional control landscape associated with multi-pulse sequences. We first introduce the optimization method for the case of unitary evolution. The computation of gradients is achieved by exploiting a time-scale separation between the two distinct interactions inside the cavity. This allows us to formulate a gradient-descent procedure to identify pulse parameters that maximize the fidelity. In a second step, we extend this framework to the dissipative regime by incorporating both Trotterization of the open-system dynamics and a controlled truncation of the underlying Hilbert space. This combined approach allows for efficient optimization in the presence of loss while keeping the computational complexity tractable. The remainder of this work is structured as follows. In Sec.~\ref{sec:idealized_scenario}, we present the optimization framework and results in the unitary case, including an analysis of the structure of optimal pulse sequences. The loss-aware optimization framework and results are presented in Sec.~\ref{sec:dissipative_dynamics}, where we separately investigate the effects of spontaneous decay of the two-level system and photon loss in the signal mode on optimal pulse sequences. Finally, Sec.~\ref{sec:conclusion} summarizes our findings and discusses possible future directions.

\section{Optimization under idealized conditions}
\label{sec:idealized_scenario}

\subsection{Modeling of Unitary Dynamics}
\label{sec:uni_modeling_of_uni_dyn}

In this subsection we introduce the description of the unitary dynamics and formulate an expression for the fidelity between the signal mode state and a target Fock state $\lvert N\rangle$ after the pulse interactions. We consider pulses of durations substantially shorter than the single-photon Rabi oscillation period, such that the atom-field interaction is negligible during the pulse interactions. The time evolution during the pulses is therefore described by the two-mode squeezing operator \cite{Krstic2024} \begin{align} \hat{U}_\mathrm{NL}(r_j,\phi_j)= \mathrm{e}^{-\mathrm{i}r_j(\mathrm{e}^{\mathrm{i}\phi_j}\hat{a}^\dagger_\mathrm{i}\hat{a}^\dagger_\mathrm{s} + \mathrm{e}^{-\mathrm{i}\phi_j}\hat{a}_\mathrm{i}\hat{a}_\mathrm{s})}\,, \end{align} where $\hat{a}_\mathrm{i}$ and $\hat{a}_\mathrm{s}$ denote the idler and signal mode annihilation operators, while $r_j$ and $\phi_j$ denote the parametric gain and phase of the $j$th pulse. In between consecutive pulses, the time evolution is described by \begin{align} \hat{U}_\mathrm{2LS}(t_j)=\mathrm{e}^{t_j\frac{\Omega}{2}(\hat{a}_\mathrm{i}\hat{\sigma}^\dagger - \hat{a}_\mathrm{i}^\dagger\hat{\sigma})}\,, \end{align} which is generated by the Jaynes-Cummings Hamiltonian \cite{Shore01071993}. Here, $\Omega$ denotes the single-photon Rabi frequency, $\hat{\sigma} = \lvert g\rangle\langle e\lvert$ is the atomic lowering operator, and $t_j$ denotes the time delay between the $j$th and $(j+1)$th pulse. Under the short-pulse approximation, the total time evolution for a sequence of $p$ pulses  can be described by a sequence of successively applied unitary operator pairs describing the
pulse-interactions and Rabi time windows as
\begin{align} \hat{U}_\mathrm{tot}(\vec{\theta}) \approx \hat{U}_\mathrm{NL}(r_p,\phi_p) \Bigl( \prod_{j=1}^{p-1} \hat{U}_\mathrm{2LS}(t_j) \hat{U}_\mathrm{NL}(r_j,\phi_j) \Bigr)\,, \label{equation:total_Unitary_Operator} \end{align} where operator pairs are ordered from left to right in descending order with respect to $j$. The parameter vector $\vec{\theta}$ collects all pulse gains $r_j$, phases $\phi_j$, and time delays $t_j$. Henceforth, composite states written as $\lvert m,n,g/e\rangle$ represent product states consisting of $m$
idler-mode excitations, $n$ signal-mode excitations, and the atomic subsystem in the ground $(g)$
or excited $(e)$ state. The state of the signal mode after the interaction with the $p$th pulse, $\hat{\rho}_\mathrm{s}(\vec{\theta})$, is obtained by taking the partial trace of the composite system density operator $\hat{\rho}_\mathrm{f} = \hat{U}_\text{tot}\lvert 0,0,g\rangle \langle 0,0,g\lvert \hat{U}_\text{tot}^\dagger$ over the atom and idler mode degrees of freedom. This yields a diagonal density operator with diagonal entries:
\begin{align}
\langle N \lvert \hat{\rho}_\mathrm{s}(\vec{\theta}) \lvert N \rangle
&=
\Bigl|
\langle N,N,g \lvert
\hat U_{\text{tot}}(\vec{\theta})
\lvert 0, 0, g \rangle
\Bigr|^2
\notag\\
&\quad+
\Bigl|
\langle N\!-\!1,N,e \lvert
\hat U_{\text{tot}}(\vec{\theta})
\lvert 0, 0, g \rangle
\Bigr|^2\,.
\label{equation:Fidelity}
\end{align}
Thus, the fidelity with respect to any target Fock state $\lvert N\rangle $
is directly given by equation \eqref{equation:Fidelity}. The next subsection will introduce the optimization scheme used to maximize the fidelity.

\subsection{Optimization of Pulse Parameters}
\label{sec:optimization_of_pulse_params}

We define the infidelity $\mathcal{I} = 1
 - \langle N\lvert \hat{\rho}_\mathrm{s}(\vec{\theta})\lvert N\rangle $ as a cost function to be minimized with respect to the pulse parameters $\vec{\theta}$. The gradient of $\mathcal{I}(\vec{\theta})$ is obtained as:
\begin{align}
    \notag \partial_{\theta_j} \mathcal{I} &= -2 \text{Re}\Bigl\{\langle N,N,g\lvert \partial_{\theta_j}\hat{U}_\text{tot}\lvert 0\rangle\langle N,N,g\lvert\hat{U}_\text{tot}\lvert 0\rangle^* \\
    &+ \langle N-1,N,e\lvert\partial_{\theta_j}\hat{U}_\text{tot}\lvert 0\rangle\langle N-1,N,e\lvert\hat{U}_\text{tot}\lvert 0\rangle^*\Bigr\}\,.
\end{align}
As $\hat{U}_\text{tot}(\vec{\theta})$ is represented as a sequence of unitary operators, the gradient components $\partial_{\theta_j}\hat{U}_\text{tot}$
 can be obtained analytically by differentiating the corresponding sub-unitary with respect to $\theta_j$
 (see App.~\ref{app:opt_unitary} for details). The optimization is performed using a gradient-descent scheme, in which the parameters are iteratively updated along the negative gradient of the cost function $\mathcal{I}(\vec{\theta})$. Our numerical computations use matrix representations of the involved operators within a truncated Fock space spanned by the states $\{\lvert n,n,g\rangle, \lvert n-1, n,e\rangle\}$ with $0\leq n \leq 60$. The maximum photon number $d=60$ for both modes is chosen to reliably capture the squeezing regimes considered in this work. Parameter updates were performed using the Adam optimizer \cite{Adam2014}, which showed markedly better performance than vanilla gradient descent. Further, the optimization was initialized from $100$ randomly chosen points in parameter space, with 
 $r_j\in [0, 15\,\text{dB}], t_j\in [0, 2\pi \Omega^{-1}]$. Throughout this work, we express
parametric gain magnitudes in dB, given by
$
r_{\mathrm{dB}} = -10 \log_{10}\!\left(e^{-2r}\right),
$ to allow for direct comparison with experimental studies of two-mode squeezed vacuum sources. The upper bound on 
 $r_j$ of $15\,\text{dB}$ was chosen to reflect the order of magnitude of experimentally achievable values \cite{Vahlbruch2016}. For each of the $100$ random initializations, the optimization was carried out independently, and the highest-fidelity solution that remained stable upon increasing the Fock-space truncation was selected and reported in the results.\\
 \indent Since numerical optimization consistently converged to pulse sequences with relative phase differences between consecutive pulses of either $0$ or $\pi$, we fixed the phase parameters to alternating values of $0$ and $\pi$ and optimized only the parameters $r_j$ and $t_j$. We allowed for $r_j\in\mathbb{R}$, including negative values, which enables effective $\pi$-phase flips relative to the fixed phase pattern and thus reduces the dimensionality of the optimization problem while preserving access to distinct phase configurations. Additional support for the optimality of the restriction to discrete phase differences between consecutive pulses is provided in the following subsection through an analytical investigation.
 
\begin{figure}[t]
	\centering
	\includegraphics[width=1\linewidth]{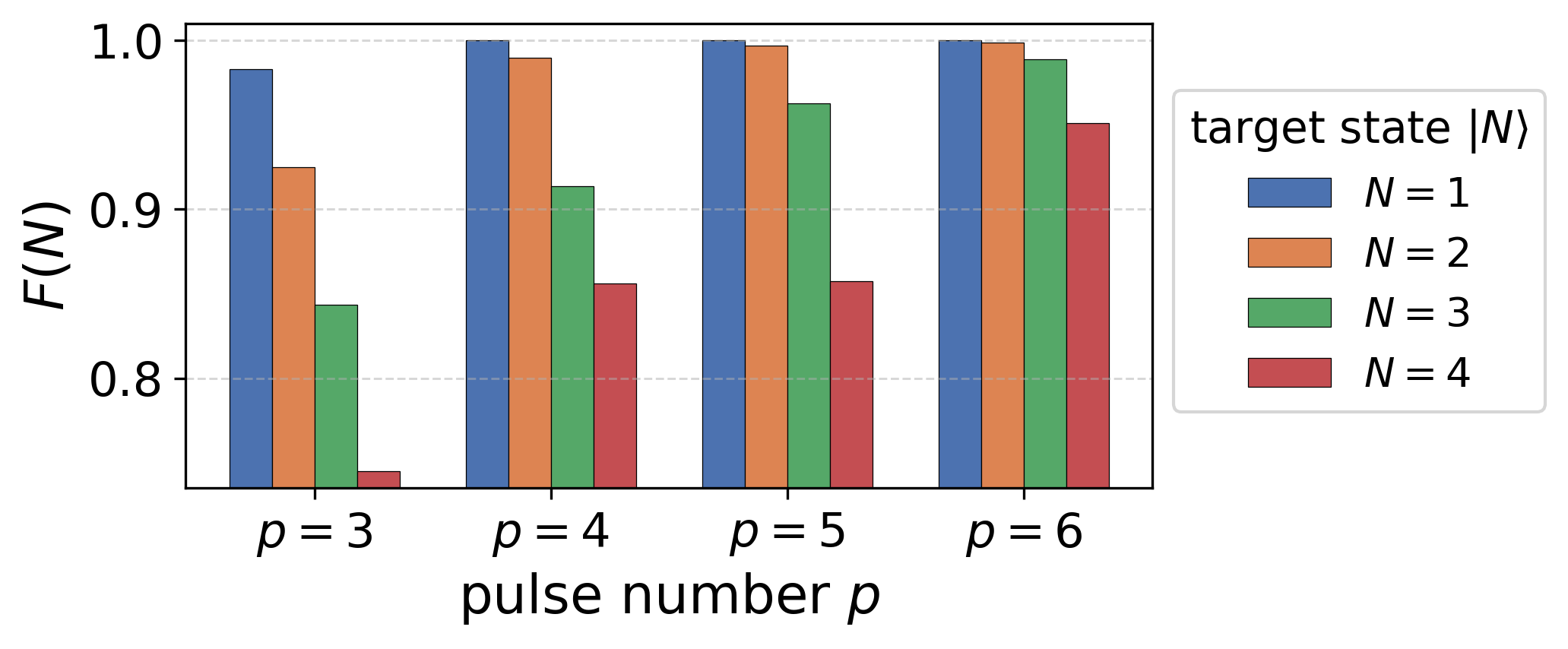}
	\caption{Maximal fidelities for target states with photon numbers $N=1,...,4$ using $p=3,...,6$ pulses assuming ideal conditions (unitary evolution).}
	\label{fig:Results_Unitary_Optimization}
\end{figure}

\subsection{Binary Phase Structure of Optimal Sequences}
\label{sec:uni_binary_phase_structure}

Here, we calculate the fidelity in an analytically solvable limit. To this end, we consider the two-pulse case and treat the second pulse as a small perturbation, such that the corresponding two-mode squeezing operator can be expanded up to first order:
\begin{align}
    \hat{U}_\text{NL}(r_2,\phi_2) = \mathbb{I} -\mathrm{i}r_2(\mathrm{e}^{\mathrm{i}\phi_2}\hat{a}_\mathrm{i}^{\dagger}\hat{a}_\mathrm{s}^{\dagger} + \mathrm{e}^{-\mathrm{i}\phi_2}\hat{a}_\mathrm{i}\hat{a}_\mathrm{s}) + \mathcal{O}(r_2^2)\,.
\end{align}
Since an analytical expression for the state of the system after the first pulse and following Rabi window is known, the fidelity of the signal mode state with respect to an arbitrary Fock state can be calculated analytically in this case (see App.~\ref{app:opt_unitary} for a detailed calculation). The result shows that a higher fidelity can be achieved using the hybrid system compared to a two-mode squeezed vacuum source, with the difference in fidelity given by:
\begin{align}
\Delta F
&= \cos(\phi_1-\phi_2)\,
   \frac{2 r_2 \tanh^{2N} r_1}{\cosh^2 r_1}\,
   \mathcal{S}(N,r_1,t)\,.
  \label{equation:perturbative_two_pulse_system}
\end{align}
From equation \eqref{equation:perturbative_two_pulse_system} it is evident that the fidelity is maximized for relative pulse phase differences of $0$ or $\pi$. The choice of phase for maximizing the fidelity in this case will depend on the sign of the function $\mathcal{S}(N,r_1,t)$ which is given in App.~\ref{app:opt_unitary}. Even though the calculation was done in a simplified scheme and we cannot necessarily extrapolate to the multi-pulse sequences used in the numerical study, it provides some physical insight. The discrete phase assumption was further validated by the numerical results.

\begin{figure*}
    \centering
    \includegraphics[width=1\linewidth]{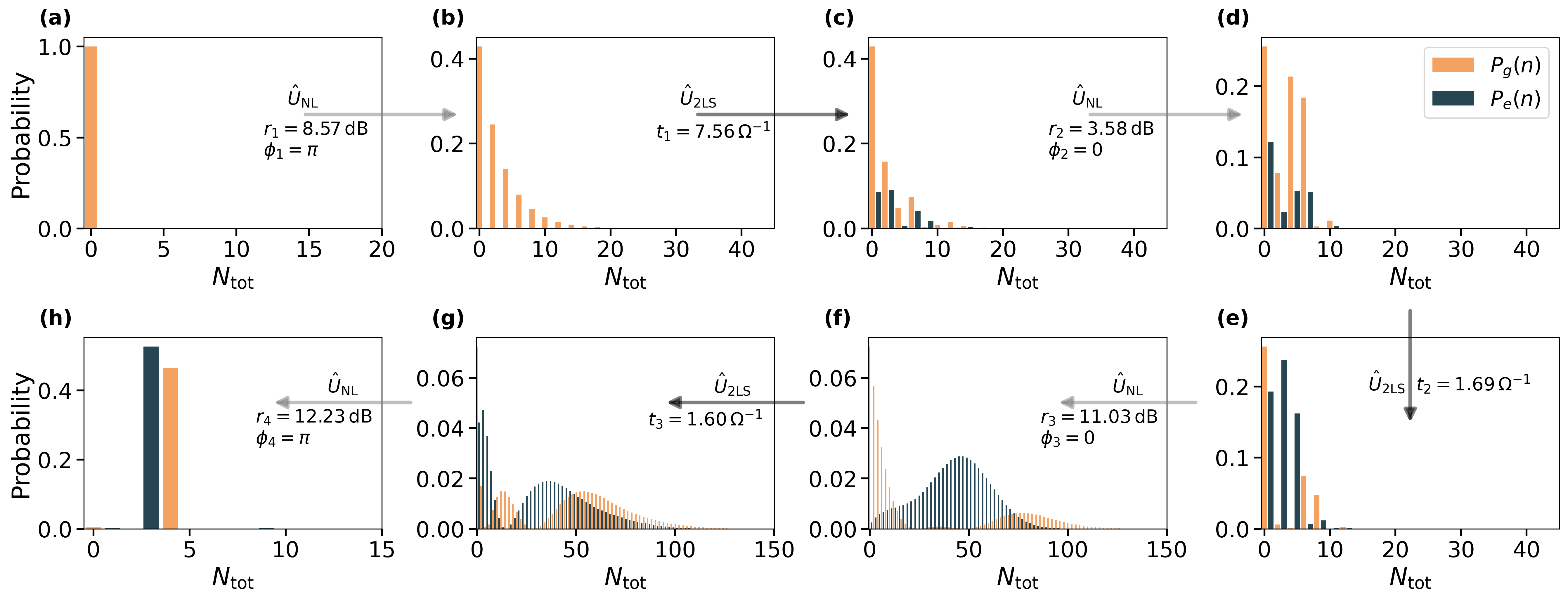}
    \caption{Composite state populations after each sub-unitary for the pulse sequence that maximizes the fidelity
    $\langle 2\lvert\hat{\rho}_\mathrm{s}\lvert 2\rangle$ in the four-pulse case. The horizontal axis shows the total photon number of the composite system. States with the atom in the ground state are represented by
$P_g(n)=\langle n,n,g|\hat{\rho}|n,n,g\rangle$ and correspond to even
total photon numbers ($N_{\mathrm{tot}}=2n$). States with the atom in
the excited state are represented by
$P_e(n)=\langle n-1,n,e|\hat{\rho}|n-1,n,e\rangle$ and correspond to odd
total photon numbers ($N_{\mathrm{tot}}=2n-1$). In the annotations, the parametric gain values and phases are shown for the squeezing transformations and the time delays are depicted for the inter-pulse delays during which the atom-field interaction takes place.}
    \label{fig:State_Transformations}
\end{figure*}

\subsection{Numerical Results}
\label{sec:uni_numerical_results}

Figure \ref{fig:Results_Unitary_Optimization} depicts the fidelities corresponding to the optimal parameter-configurations obtained from the optimization using pulse sequences of $3$ to $6$ pulses while targeting low-photon-number states $\lvert N\rangle$ with ${N \in \{1,2,3,4\}}$. The 3-pulse fidelities correspond to the parameter-configurations presented in \cite{Krstic2024} and were reproduced using our gradient-descent optimization. While near-deterministic state preparation was already possible for the single-photon state using three pulses, multi-pulse sequences of four or more pulses introduce more control degrees of freedom and can enhance state preparation through more complex interference patterns. As a result, Fock states of two and three photons can be prepared with near-unity probability under ideal conditions using four- to six-pulse sequences, and the four-photon Fock state can be prepared with $95\,\%$ fidelity using a six-pulse sequence. Interestingly, the highest-fidelity configurations show some relative phase differences of $\Delta \phi = 0$ between consecutive pulses, which differs from the initial configuration (the optimal parameter values are tabulated in App.~\ref{app:opt_unitary}). To illustrate how the combination of squeezing and atom-field interaction can produce a Fock state in the signal mode, Fig.~\ref{fig:State_Transformations} shows how the composite system evolves after each sub-unitary for the pulse parameters that maximize the fidelity to the $\lvert N=2\rangle$ state in a four-pulse setting. In the following subsection, we discuss this exemplary preparation sequence in detail and identify the relevant mechanisms leading to near-deterministic preparation. 

\begin{figure*}
    \centering
    \includegraphics[width=1\linewidth]{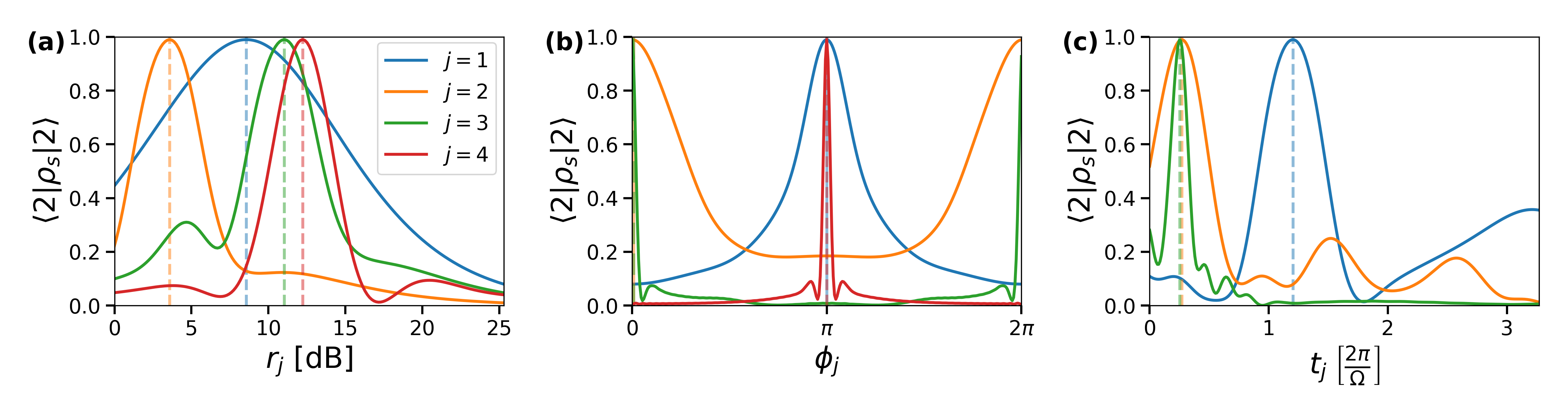}
    \caption{Fidelity $\langle 2\lvert\hat{\rho}_\mathrm{s}(\vec{\theta})\lvert 2\rangle$ for the optimal point in parameter space $\vec{\theta}$ in the unitary four-pulse setting. (a) depicts the dependency on the parametric gain magnitudes $r_j$, (b) that on the phases $\phi_j$ and (c) shows how the time-delays $t_j$ influence the fidelity. In all subfigures the vertical dashed lines represent the value of the respective optimized parameter $\theta_j$.}
    \label{fig:Optimization_Landscape_three_subfigs}
\end{figure*}

\subsection{Analysis of the Preparation Process}
\label{sec:uni_preparation_process}

Initially, the system is in the vacuum state $\lvert 0,0,g\rangle$ (Fig.~\hyperref[fig:State_Transformations]{\ref*{fig:State_Transformations}a}). After the first pulse, a two-mode squeezed vacuum state is generated (Fig.~\hyperref[fig:State_Transformations]{\ref*{fig:State_Transformations}b}), resulting in a distribution over composite states of the form $\lvert n,n,g\rangle$, i.e. states with equal photon numbers in the signal and idler modes. The subsequent Jaynes-Cummings evolution couples these even-total-photon-number states to states of the form $\lvert n-1,n,e\rangle$, thereby transferring population into the odd-total-photon-number sector (Fig.~\hyperref[fig:State_Transformations]{\ref*{fig:State_Transformations}c}).\\
\indent The second and third pulse and the intervening Rabi-interaction period (Figs.~\hyperref[fig:State_Transformations]{\ref*{fig:State_Transformations}d}--\hyperref[fig:State_Transformations]{\ref*{fig:State_Transformations}f}) then gradually drive the system into a nearly perfect superposition of the two squeezed composite states
$\hat{U}_{\mathrm{NL}}(r_{\mathrm{eff}}, \phi_{\mathrm{eff}})\lvert 2,2,g\rangle$
and
$\hat{U}_{\mathrm{NL}}(r_{\mathrm{eff}}, \phi_{\mathrm{eff}})\lvert 1,2,e\rangle$ with effective squeezing parameters $r_\mathrm{eff}, \phi_\mathrm{eff}$ (Fig.~\hyperref[fig:State_Transformations]{\ref*{fig:State_Transformations}g}). Finally, a squeezing pulse with parameters $r_4 = r_{\mathrm{eff}}$ and $\phi_4 = \phi_{\mathrm{eff}} + \pi$ is applied. Since
$
\hat{U}_{\mathrm{NL}}(r_\mathrm{eff},\phi_\mathrm{eff})\hat{U}_{\mathrm{NL}}(r_\mathrm{eff},\phi_\mathrm{eff}+\pi) = \mathbb{I},
$
this pulse reverses the effective previous squeezing transformation and projects the state onto a superposition of the two composite states $\lvert 2,2,g\rangle$ with $N_\mathrm{tot}=4$ and $\lvert 1,2,e\rangle$  with $N_\mathrm{tot}=3$ (Fig.~\hyperref[fig:State_Transformations]{\ref*{fig:State_Transformations}h}). These are precisely the states that contribute to a signal-mode occupation of two photons. Effectively, the Rabi oscillation periods act as a photon-number-dependent tuning mechanism, coupling the $\lvert n,n,g\rangle$-sector to the $\lvert n-1,n,e\rangle$-sector. Combined with the rich dynamics of two-mode squeezing, this enables the transformation from the two-mode squeezed vacuum state in Fig.~\hyperref[fig:State_Transformations]{\ref*{fig:State_Transformations}b} to the superposition of squeezed composite states in Fig.~\hyperref[fig:State_Transformations]{\ref*{fig:State_Transformations}g}.
\subsection{Optimization Landscape}
To provide insight into the optimization landscape, Fig.~\ref{fig:Optimization_Landscape_three_subfigs} depicts how the fidelity changes if one of the parameters is varied while keeping all others fixed at the values of the optimal pulse sequence from Fig.~\ref{fig:State_Transformations}. Note that the optimization landscape generally changes from point to point in parameter space as any one of the parameters is varied. Hence
these one-dimensional slices can only provide a partial view into the structure of the 11-dimensional
parameter space. Nevertheless, in a region close to the investigated point $\vec{\theta}$, the gradient $\nabla F(\vec{\theta})$
can be estimated from observing these one-dimensional cuts. The slices of the fidelity landscape exhibit sharp maxima, particularly with respect to the time-delay and phase parameters, whereas the gain magnitudes produce smoother variations. The presence of these steep variations explains the superior performance of the Adam optimizer and the necessity of using multiple random initializations to reliably locate high-fidelity solutions. We show the results for each initialization of the four-pulse optimization from which we obtained the parameter configuration corresponding to Fig.~\ref{fig:State_Transformations} and Fig.~\ref{fig:Optimization_Landscape_three_subfigs} in App.~\ref{app:opt_unitary}. 

\section{Optimization under loss}
\label{sec:dissipative_dynamics}

\subsection{Modeling of Dissipative Dynamics}
\label{sec:modeling_of_diss_dyn}

We describe the dissipative dynamics using a channel representation based on the Lindblad master equation (LME) \cite{Manzano2019}. During time intervals of pump-pulse interactions with the nonlinear crystal, we neglect losses. This approximation remains valid as long as decay times of interest are much longer than the pump pulse lengths. The resulting dynamics are expressed as a sequence of quantum channels acting on the density operator, alternating between unitary nonlinear pulse channels $\mathcal{E}_\text{NL}(r_j, \phi_j)$ and dissipative channels $\mathcal{E}_\text{2LS}(t_j)$ generated by a Lindbladian during the inter-pulse delays,
\begin{align}
\hat{\rho}_\mathrm{f}  \approx  \mathcal{E}_\text{NL}(r_p,\phi_p)\circ \Bigl(\bigcirc_{j=1}^{p-1} [\mathcal{E}_\text{2LS}(t_j)\circ\mathcal{E}_\text{NL}(r_j,\phi_j)]\Bigr)(\hat{\rho}_{\text{vac}})\,,
\label{equation:Lindbladian_Channel_Representation_of_Dynamics}
\end{align}
analogous to the representation of the dynamics in the unitary case in Eq.~\eqref{equation:total_Unitary_Operator}. Here $\circ$ denotes composition of maps, so the rightmost channel acts first on $\hat{\rho}_{\mathrm{vac}}=\lvert 0,0,g\rangle\langle0,0,g\lvert$. The maps are written from left to right in descending order of $j$. Considering the hybrid system as an open system, the fidelity to a Fock state $\lvert N\rangle$ is obtained as the sum over all diagonal entries of the final composite state density operator that correspond to the signal mode being occupied by $N$ photons, including those populated through dissipative processes:
\begin{align}
 \langle N\lvert \hat{\rho}_\mathrm{s}(\vec{\theta})\lvert N\rangle  &= \sum\limits_{a \in \{g,e\}} \sum\limits_{i =0}^\infty\langle i, N, a\lvert \hat{\rho}_\mathrm{f}(\vec{\theta})\lvert i,N,a\rangle\,.
\end{align}
The gradient of the infidelity with respect to the control parameters $\nabla_{\vec{\theta}}\, \mathcal{I}(\vec{\theta})$ is therefore given by the corresponding matrix elements of the gradient of the final composite state density operator. Gradient components of $\hat{\rho}_\mathrm{f}(\vec{\theta})$ with respect to the control parameters are obtained by differentiating the individual quantum channels in the channel decomposition \eqref{equation:Lindbladian_Channel_Representation_of_Dynamics}. See App.~\ref{app:opt_loss} for the explicit calculation of gradients. \\
\indent Dissipation substantially increases computational demands by coupling the system to a larger Hilbert space than would be sufficient for the unitary case. Furthermore, a Liouville-space representation requires $D^2\times D^2$ matrices for a total truncated Fock space of dimension $D$. To make the optimization tractable, we employ a Trotterized decomposition \cite{Blanes_Casas_Murua_2024, PhysRevX.11.011020} of the dissipative channels, separating coherent and incoherent contributions at each timestep using a symmetric splitting scheme with controllable error. This Trotterization is computationally more efficient since it avoids an explicit Liouville-space representation and instead allows the dynamics to be evaluated directly in the physical Hilbert space. Together with a Hilbert-space truncation tailored to the considered decay mechanism, this approach enables efficient simulation and optimization of the dissipative dynamics. To consider the system dynamics under more realistic conditions, we investigate the influence of spontaneous emission from the 2LS and photon loss in the signal mode on the optimal pulse parameters.

\begin{figure}
    \centering
    \includegraphics[width=0.9\linewidth]{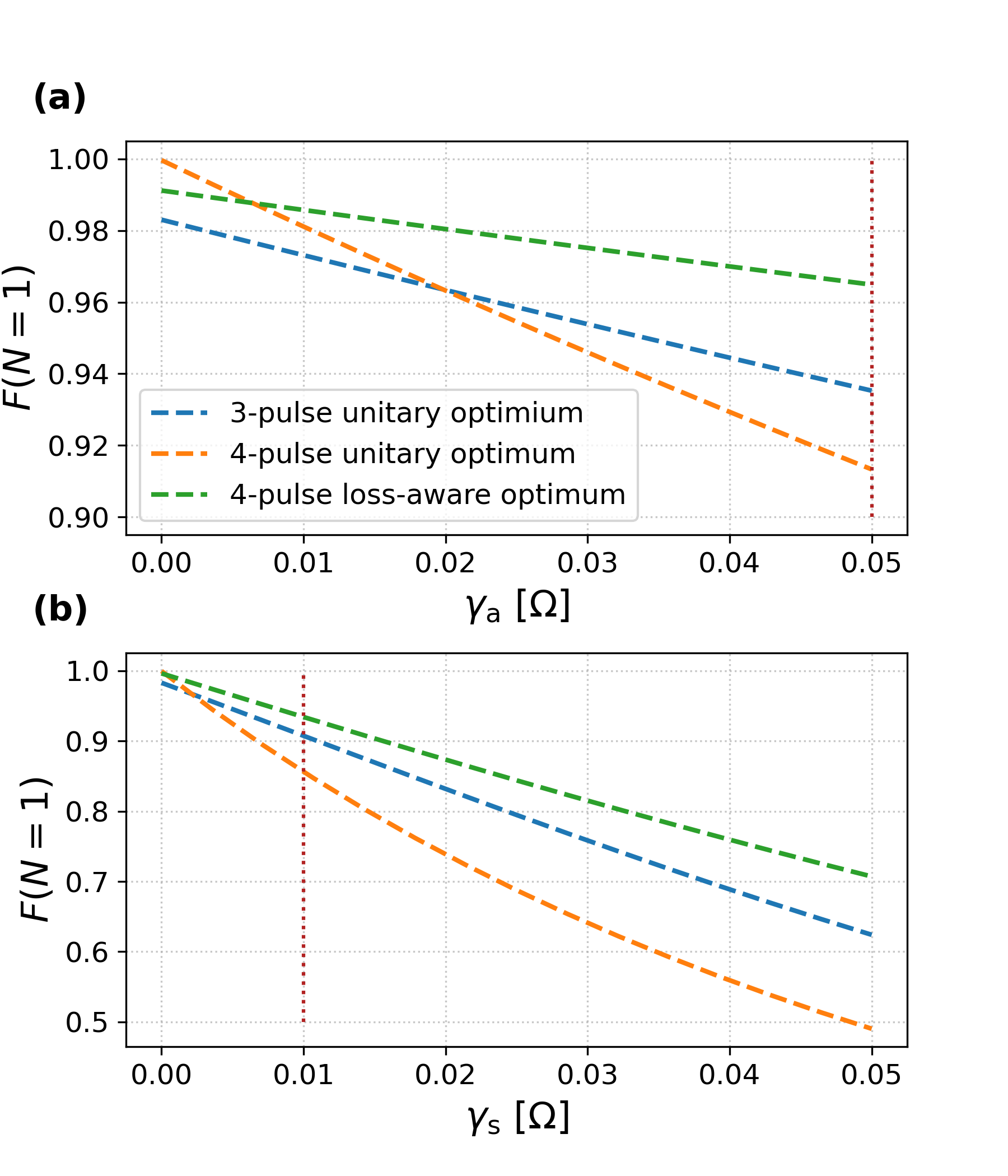}
    \caption{Comparison of fidelity to the single photon Fock state for three different parameter-optima depending on the dissipation rate for atom decay (a) and photon loss in the signal mode (b). The dissipation rates at which the Lindbladian optimization was performed to obtain the robust parameters are indicated by the vertical dashed lines in red.}
\label{fig:dissipation_comparison}
\end{figure}

The quantum channel describing atom decay, i.e. the hybrid system spontaneously emitting a photon outside the cavity through the 2LS, can be calculated for a time step $\Delta t$ by applying the following transformations to the respective subsectors of the entire composite system density operator
$\hat{\rho}(t)$ that correspond to the atomic degrees of freedom:
\begin{align}
    \begin{pmatrix}
\rho_{gg}(t')  \\
\rho_{eg}(t') \\
\rho_{ge}(t')  \\
\rho_{ee}(t') 
\end{pmatrix} = \begin{pmatrix}
    1 & 0 & 0 & 1-\mathrm{e}^{-\gamma_\mathrm{a}\Delta t} \\
    0 & \mathrm{e}^{-\frac{\gamma_\mathrm{a}}{2}\Delta t} & 0 & 0 \\
    0 & 0 & \mathrm{e}^{-\frac{\gamma_\mathrm{a}}{2}\Delta t} & 0 \\
    0 & 0 & 0 & \mathrm{e}^{-\gamma_\mathrm{a}\Delta t}\end{pmatrix}
    \begin{pmatrix}
        \rho_{gg}(t)  \\
\rho_{eg}(t) \\
\rho_{ge}(t)  \\
\rho_{ee}(t) 
    \end{pmatrix}
\label{equation:pure_atom_decay_channel}
\end{align}
with $\gamma_\mathrm{a}$ being the atomic decay rate.\\
\indent For photon loss, an analytical solution to the corresponding dissipative channel exists using time-dependent Kraus operators \cite{Fan2019}. It can be expressed as an infinite time-dependent series, which reduces to two leading terms in the short-time limit. For a time step $\Delta t\ll \,\Omega^{-1}$, we can approximate the quantum map as
\begin{align}
    \hat{\rho}(t+\Delta t) \approx \hat{M}_0\hat{\rho}(t)\hat{M}^\dagger_0 + \hat{M}_1\hat{\rho}(t)\hat{M}^\dagger_1\,, 
\label{equation:pure_photon_loss_channel}
\end{align}
with $\hat{M}_0 = \mathrm{e}^{-\frac{\gamma_\mathrm{s}}{2}\Delta t \hat{a}_\mathrm{s}^\dagger\hat{a}_\mathrm{s}}$ and $\hat{M}_1 = \sqrt{1-\mathrm{e}^{\gamma_\mathrm{s}\Delta t}} \,
\mathrm{e}^{-\frac{\gamma_\mathrm{s}}{2}\Delta t \hat{a}_\mathrm{s}^\dagger\hat{a}_\mathrm{s}}
\hat{a}_\mathrm{s}\,.$ Here $\gamma_\mathrm{s}$ describes the photon loss rate. 
The optimization was carried out using the above transformations of the density operator for atom decay and photon loss, respectively, in a Trotterization scheme while using a truncated Fock-space basis, retaining only those states that acquire non-negligible occupation during the relevant interaction times (see App.~\ref{app:opt_loss} for details).

\subsection{Numerical Results}
\label{sec:loss_numerical_results}

\subsubsection{Single Photon Target State}
\label{sec:loss_1phot}

\begin{figure}
    \centering
    \includegraphics[width=0.95\linewidth]{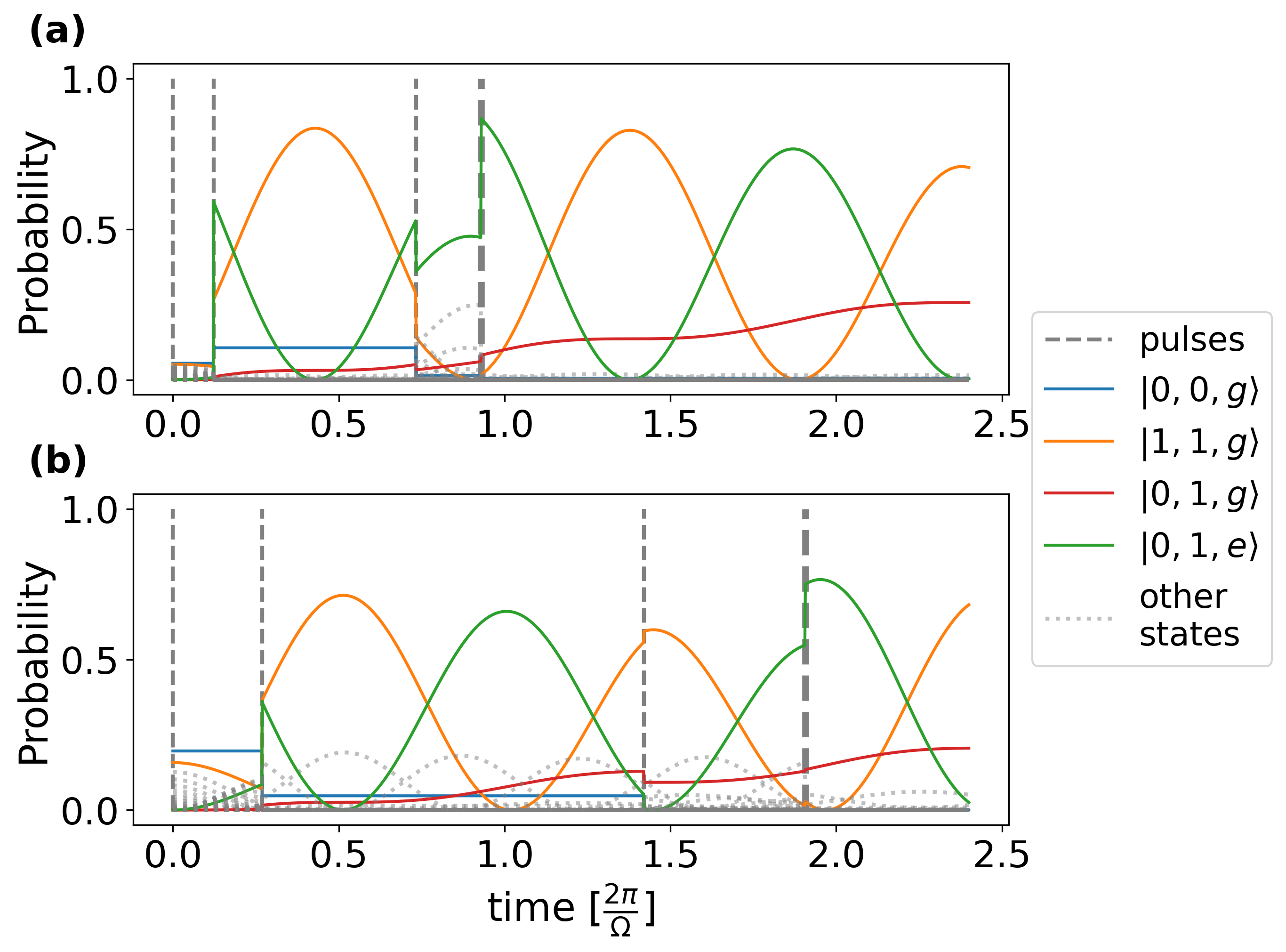}
    \caption{Time evolution of composite state populations for atom decay with a decay rate of $\gamma_\mathrm{a} = 0.05\,\Omega$. (a) shows the dynamics corresponding to the parameter configuration $\vec{\theta}_\text{LME}$ obtained from the loss-aware optimization, yielding a fidelity at the end of the preparation protocol of $\langle 1\lvert \hat{\rho}_\mathrm{s}(\vec{\theta}_\text{LME})\lvert 1\rangle = 0.965\,.$ (b) shows the dynamics corresponding to the parameter configuration $\vec{\theta}_\text{U}$ obtained from the unitary optimization with final fidelity  $\langle 1\lvert \hat{\rho}_\mathrm{s}(\vec{\theta}_\text{U})\lvert 1\rangle = 0.913\,.$ States contributing to the fidelity $\langle 1\lvert\hat{\rho}_\mathrm{s}\lvert 1\rangle$, together with the vacuum state are shown explicitly, while all other states are indicated by gray dotted lines. The times when the pulses are applied are indicated by the vertical dashed lines. The end of the preparation protocol, corresponding to the fourth pulse, is highlighted by a thicker vertical dashed line.}
    \label{fig:Dissipative_Dynamics_Atom_Decay_gamma=0.05}
\end{figure}

\begin{figure}
    \centering
    \includegraphics[width=0.95\linewidth]{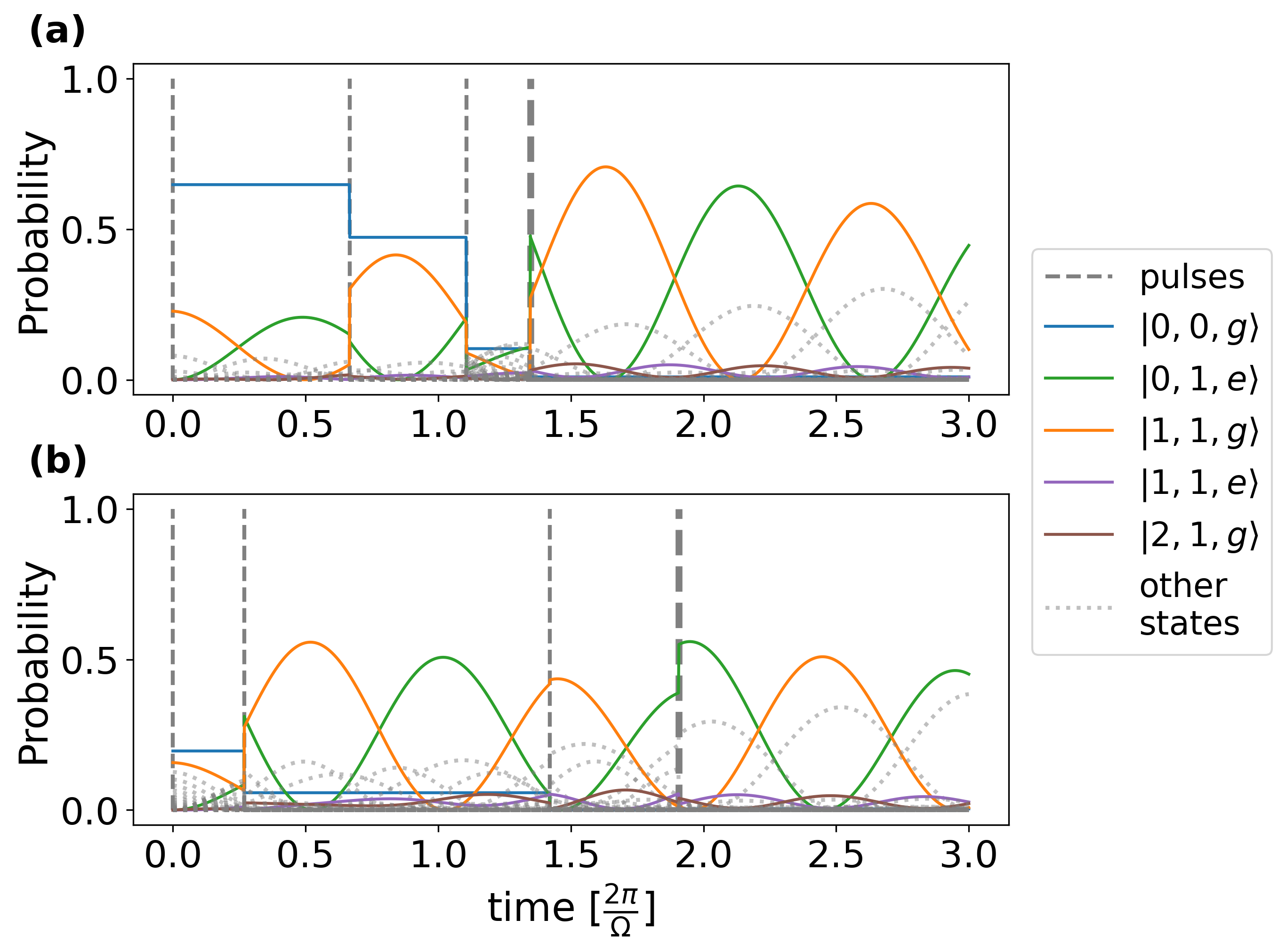}
    \caption{Time evolution of composite state populations for photon loss with a decay rate of $\gamma_\mathrm{s} = 0.03\,\Omega$. (a) shows the dynamics corresponding to the parameter configuration $\vec{\theta}_\text{LME}$ obtained from the loss-aware optimization. The fidelity after the fourth pulse there amounts to $\langle 1\lvert \hat{\rho}_\mathrm{s}(\vec{\theta}_\text{LME})\lvert 1\rangle =  0.815\,.$ (b) shows the dynamics corresponding to the parameter configuration $\vec{\theta}_\text{U}$ obtained from the unitary optimization with final fidelity  $\langle 1\lvert \hat{\rho}_\mathrm{s}(\vec{\theta}_\text{U})\lvert 1\rangle = 0.641\,.$ Only the main contributing states to the fidelity $\langle 1\lvert \hat{\rho}_\mathrm{s}\lvert 1\rangle$ as well as the vacuum state $\lvert 0,0,g\rangle$ are highlighted, all others are shown as dotted lines.}
    \label{fig:Dissipative_Dynamics_photon_loss_gamma=0.03}
\end{figure}

We present in this section the results from the optimization of the single-photon Fock state in a four-pulse setup under atom decay and photon loss, respectively. Results on the influence of photon loss on the preparation of the $\lvert N=2\rangle$ state are shown in Sec.~\ref{sec:loss_2phot}. Fig.~\ref{fig:dissipation_comparison} shows the fidelity corresponding to three different parameter configurations: (i) the three-pulse configuration reported in \cite{Krstic2024}, (ii) the four-pulse configuration obtained via optimization of the idealized unitary state preparation, and (iii) the parameter set derived from the loss-aware optimization, which was carried out for a fixed value of the dissipation rate. The fidelities corresponding to the three parameter sets are computed under the same dissipative conditions across a range of dissipation rates for atom decay and photon loss, respectively. This allows for a direct comparison of their robustness with respect to increasing dissipation.\\
\indent The loss-aware optima exhibit enhanced robustness against the respective decay mechanisms, with higher loss rates accentuating the advantage of the parameters optimized under dissipation over those optimal in the unitary case. Interestingly, the highest fidelity obtained from the unitary four-pulse optimization suffers substantially under loss, such that the three-pulse configuration already performs better after a certain magnitude of the dissipation rate. This shows that optimizing the preparation process in the unitary case with more pulses does not suffice to find better solutions in more realistic settings. The dissipation-aware optimization hence becomes a necessary tool to discover parameter configurations for which Fock state preparation is more robust to loss.\\
\indent We identify two main factors contributing to the enhanced robustness of the state preparation success for the parameters obtained from the loss-aware optimization. Figures \ref{fig:Dissipative_Dynamics_Atom_Decay_gamma=0.05} and \ref{fig:Dissipative_Dynamics_photon_loss_gamma=0.03} show the evolution of composite state populations for the optimal configurations under atomic decay and photon loss, respectively, with the dynamics of the unitary optimum under the respective decay mechanism also shown for comparison in both cases. 

Especially for photon loss, which is far more detrimental to the fidelity than atom decay, dissipation-aware optimization yields substantially better parameters than the unitary optimization, resulting in an absolute fidelity improvement of $17.4\,\%$.
This robustness can be attributed to (i) the pulse protocol's markedly shorter total interaction time and (ii) the concentration of population into a minimal set of composite states during relevant inter-pulse delays  (see comparison of the time-delay between the second and third pulse in Figures \ref{fig:Dissipative_Dynamics_Atom_Decay_gamma=0.05} and \ref{fig:Dissipative_Dynamics_photon_loss_gamma=0.03}, respectively). In the robust solution, mainly the vacuum state and a small set of states that directly contribute to the target fidelity are populated, which reduces coupling to states in the loss sector. In contrast, additional loss-sensitive composite states become excited in the dynamics corresponding to the unitary optimum. This mechanism is effective for both atomic decay and photon loss.\\
\indent Another feature of the more robust pulse-sequence against photon loss is that the Lindbladian solution requires lower parametric gain than the three-pulse configuration from \cite{Krstic2024}, making it the far more favorable choice as larger squeezing is harder to produce experimentally.

\subsubsection{Two Photon Target State}
\label{sec:loss_2phot}

\begin{figure}
    \centering
    \includegraphics[width=0.9\linewidth]{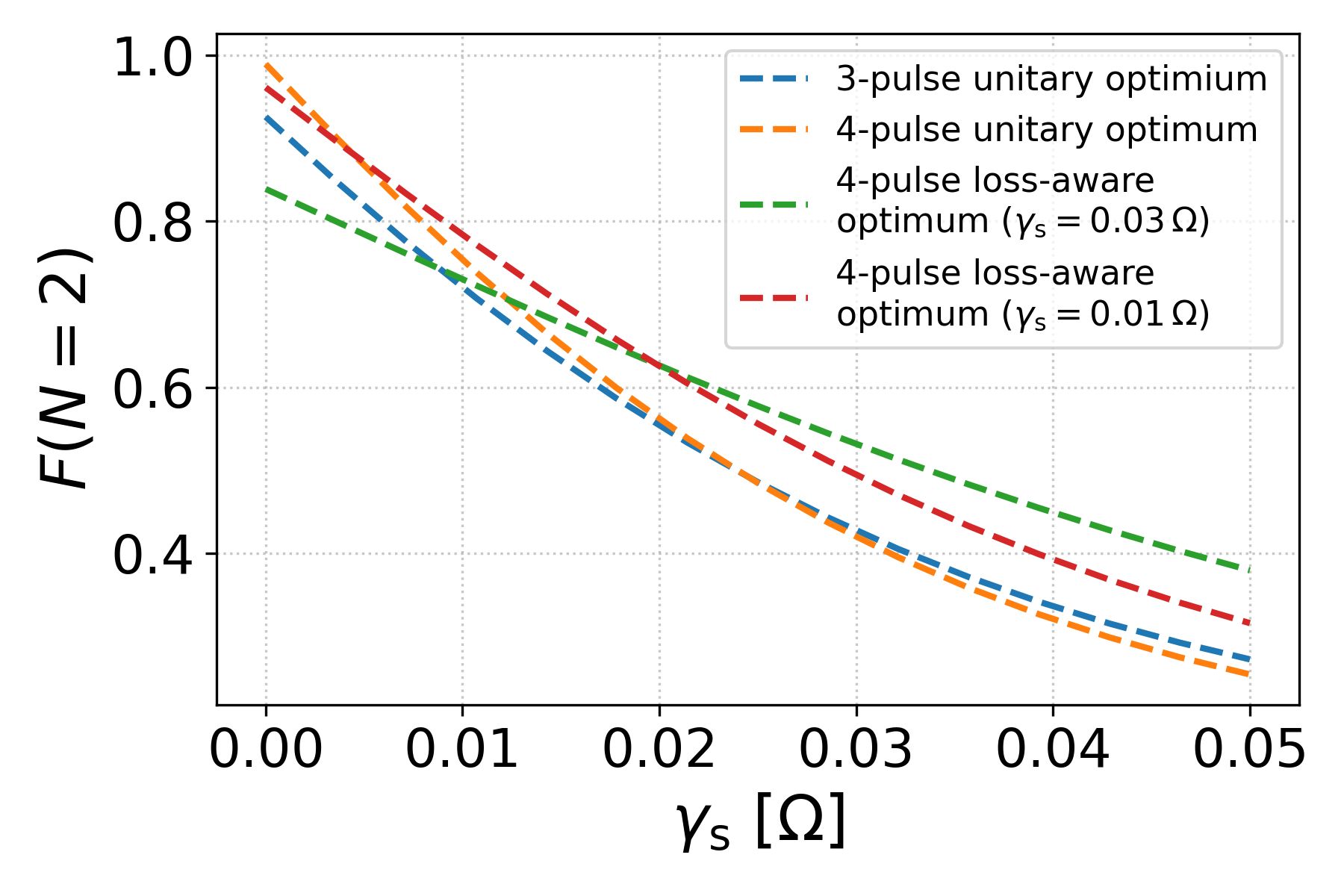}
    \caption{Comparison of fidelity to the two-photon Fock state for four different parameter-optima depending on the photon loss rate in the signal mode.}
    \label{fig:N2p4_LME_optimum_photon_loss_dissipation_rate_comparison}
\end{figure}

For the target Fock state with two photons we ran the optimization once for a photon loss rate of $\gamma_\mathrm{s}=0.01\,\Omega$ and once for $\gamma_\mathrm{s}=0.03\,\Omega$ to investigate the influence of the dissipation rate used during optimization on the robustness of the obtained optimum. Figure \ref{fig:N2p4_LME_optimum_photon_loss_dissipation_rate_comparison} shows the fidelities as a function of the photon-loss rate for the optimized parameter sets, along with two reference solutions: the three-pulse optimum from \cite{Krstic2024} and the optimal parameter set obtained in the unitary four-pulse case. Generally, photon loss has a stronger negative effect on the preparation than for the single-photon state.\\
\indent The comparison between the different pulse sequences shown in Fig.~\ref{fig:N2p4_LME_optimum_photon_loss_dissipation_rate_comparison} reveals that distinct parameter constellations are optimal in different regimes of the photon-loss rate. While the optimum obtained in the unitary setting still yields the highest fidelities for very small dissipation rates, in the intermediate regime ($0.008\,\Omega \leq \gamma_s \leq 0.02\,\Omega$), the parameter set optimized for $\gamma_s = 0.01\,\Omega$ provides the best performance. For larger loss rates ($\gamma_s \geq  0.02\,\Omega$), the parameters obtained from the optimization performed at $\gamma_s = 0.03\,\Omega$ become the best-performing solution.\\
\indent Consistently, the unitary four-pulse optimum outperforms both loss-aware solutions in the absence of photon loss, indicating that pulse sequences optimized for dissipative conditions are generally not optimal in the lossless regime. These results demonstrate that the optimization can be tailored to the experimentally relevant dissipation regime in order to maximize the achievable fidelity under the corresponding operating conditions. Interestingly, for sufficiently large loss rates, the best-performing pulse sequence is one that attains a unitary fidelity of only slightly above $80\%$. The fact that a pulse sequence exhibiting relatively modest performance in the absence of loss becomes optimal in the strongly dissipative regime highlights the impact of photon loss on the optimization landscape and shows that dissipation modifies the optimal state-preparation process.

\section{Conclusions}
\label{sec:conclusion}

In this work, we have developed an optimization framework for the preparation of photonic Fock states in a hybrid quantum optical system. Under the assumption of unitary dynamics, this approach enables efficient optimization of multi-pulse sequences and demonstrates that near-unity generation probabilities can be systematically approached for low-photon-number Fock states. One finding from our numerical results is the emergence of a binary phase structure in the optimal pulse sequences, with relative phases restricted to values of $0$ or $\pi$. We provide analytical evidence for this behavior in a simplified setting, supporting the observation that the control landscape effectively enforces this phase quantization. A more general theoretical formulation of this principle, which extends beyond the present model, remains an interesting direction for future work. Further extensions could also include the investigation of continuously driven pump fields instead of discrete pulse sequences, which may further improve state preparation performance. By allowing for arbitrarily large parametric gains and interaction times, a future theoretical study could also systematically explore up to which photon number $N$ deterministic Fock-state preparation remains feasible in this system.\\
\indent By expanding the framework to open-system dynamics,
we demonstrated that incorporating dissipation directly
into the optimization is essential for identifying the best-performing protocols under realistic experimental conditions. The resulting sequences enable state preparation success with enhanced robustness by minimizing interaction times and suppressing population transfer into dissipative channels.\\
\indent Taken together, our results establish a versatile optimization framework for the preparation of photonic Fock states in hybrid quantum optical systems. The resulting multi-pulse control protocols, together with the identified phase constraints and dissipation-aware optimization, enable robust high-fidelity Fock-state preparation with improved resilience against decoherence. These results provide a concrete route toward scalable photonic quantum state engineering with potential impact on future quantum technologies.
\section*{Acknowledgments}
M.G. acknowledges support by the Federal Ministry of Research, Technology and Space (BMFTR) under project BeRyQC. S.S. acknowledges funding by the Nexus program of the Carl-Zeiss-Stiftung (project MetaNN, project ID P2022-
04-018).
This research is supported by funding from the German Research Foundation (DFG) under the project identifiers Grant No.\ 398816777-SFB 1375 (NOA project A7), and Grant No.\ 550495627-FOR 5919 (MLCQS), and the Carl-Zeiss-Stiftung (CZS Center QPhoton).


Large Language Models (ChatGPT and Claude) were employed to enhance the manuscript’s clarity and style, including the shortening and restructuring of parts of the text. The authors reviewed and edited all outputs. All scientific content and conclusions were generated by the authors, who maintain full  accountability for the final text.


The data that support the findings of this article will be publicly available after acceptance.

\bibliography{bibliography}

\clearpage 

\appendix
\section{Optimization under Idealized Conditions}
\label{app:opt_unitary}
\subsection{Gradients}
The partial derivatives required for the computation of the gradient of $\mathcal{I}(\vec{\theta})$ are given by
\begin{align}
    \partial_r \hat{U}_\text{NL} &= -\mathrm{i}(\mathrm{e}^{\mathrm{i}\phi}\hat{a}_\mathrm{i}^{\dagger}\hat{a}_\mathrm{s}^{\dagger}+\mathrm{e}^{-\mathrm{i}\phi}\hat{a}_\mathrm{i}\hat{a}_\mathrm{s}) \hat{U}_\text{NL}\,,\\
    \partial_\phi \hat{U}_\text{NL} &= r(\mathrm{e}^{\mathrm{i}\phi}\hat{a}_\mathrm{i}^{\dagger}\hat{a}_\mathrm{s}^{\dagger}-\mathrm{e}^{-\mathrm{i}\phi}\hat{a}_\mathrm{i}\hat{a}_\mathrm{s}) \hat{U}_\text{NL}\,,\\
    \partial_t \hat{U}_\text{2LS} &= \frac{\Omega}{2} (\hat{a}_\mathrm{i}\hat{\sigma}^{\dagger}-\hat{a}_\mathrm{i}^{\dagger}\hat{\sigma} )\hat{U}_\text{2LS}\,.
\end{align}
\subsection{Phase Quantization}

\subsubsection{Perturbative Treatment of the Two-Pulse Scenario}

We consider the two-pulse sequence
\begin{align}
\hat{U} = \hat{U}_\text{NL}(r_2,\phi_2)\hat{U}_\text{2LS}(t)\hat{U}_\text{NL}(r_1,\phi_1)
\end{align}
acting on the initial state $\lvert 0,0,g\rangle$. After the first pulse, the system is prepared in a two-mode squeezed vacuum state,
\begin{align}
\hat{U}_\text{NL}(r_1,\phi_1)\ket{0,0,g} = \sum_{n=0}^\infty \frac{
\left[-\mathrm{i}\mathrm{e}^{\mathrm{i}\phi_1}\tanh r_1\right]^n
}{\cosh r_1} \ket{n,n,g}.
\end{align}
The subsequent two-level system evolution maps each equal-photon-number composite state according to
\begin{align}
\ket{n,n,g} \rightarrow
\cos\!\Bigl(\sqrt{n}\tfrac{\Omega t}{2}\Bigr)\ket{n,n,g}
+ \sin\!\Bigl(\sqrt{n}\tfrac{\Omega t}{2}\Bigr)\ket{n-1,n,e},
\end{align}
resulting in a superposition of 2LS ground-state and excited-state contributions. Expanding the operator that represents a second pulse of small squeezing to first order in $r_2$ then yields
\begin{align}
\hat{U}\ket{0,0,g}
\approx \frac{1}{\cosh r_1}\sum_{n=0}^\infty c_n \Bigl[\mathcal{A}_n^{(g)} + \mathcal{A}_n^{(e)}\Bigr],
\end{align}
with $c_n = [-\mathrm{i}\mathrm{e}^{\mathrm{i}\phi_1}\tanh r_1]^n$ and
\begin{align}
\mathcal{A}_n^{(g)} &= \cos\!\Bigl(\sqrt{n}\tfrac{\Omega t}{2}\Bigr)
\Bigl[\ket{n,n,g} \nonumber \\
&\quad - \mathrm{i}r_2 \mathrm{e}^{\mathrm{i}\phi_2}(n+1)\ket{n+1,n+1,g} \nonumber \\
&\quad - \mathrm{i}r_2 \mathrm{e}^{-\mathrm{i}\phi_2}n\ket{n-1,n-1,g}\Bigr], \\
\mathcal{A}_n^{(e)} &= \sin\!\Bigl(\sqrt{n}\tfrac{\Omega t}{2}\Bigr)
\Bigl[\ket{n-1,n,e} \nonumber \\
&\quad - \mathrm{i}r_2 \mathrm{e}^{\mathrm{i}\phi_2}\sqrt{n(n+1)}\ket{n,n+1,e} \nonumber \\
&\quad - \mathrm{i}r_2 \mathrm{e}^{-\mathrm{i}\phi_2}\sqrt{n(n-1)}\ket{n-2,n-1,e}\Bigr].
\end{align}

\subsubsection{Relevant Transition Amplitudes}

Projecting onto the relevant states $\ket{N,N,g}$ and $\ket{N-1,N,e}$ yields
\begin{align}
A_g &= \frac{1}{\cosh r_1}\Bigl[
c_N C_N 
- \mathrm{i}r_2 N \mathrm{e}^{\mathrm{i}\phi_2} c_{N-1} C_{N-1} \nonumber \\
&\qquad\qquad
- \mathrm{i}r_2 (N+1)\mathrm{e}^{-\mathrm{i}\phi_2} c_{N+1} C_{N+1}
\Bigr], \\
A_e &= \frac{1}{\cosh r_1}\Bigl[
c_N S_N 
- \mathrm{i}r_2 \sqrt{N(N-1)} \mathrm{e}^{\mathrm{i}\phi_2} c_{N-1} S_{N-1} \nonumber \\
&\qquad\qquad
- \mathrm{i}r_2 \sqrt{N(N+1)} \mathrm{e}^{-\mathrm{i}\phi_2} c_{N+1} S_{N+1}
\Bigr]
\end{align}
where
\begin{align}
C_n = \cos\!\Bigl(\sqrt{n}\tfrac{\Omega t}{2}\Bigr), 
\qquad
S_n = \sin\!\Bigl(\sqrt{n}\tfrac{\Omega t}{2}\Bigr).
\end{align}

\subsubsection{Fidelity to First Order in $r_2$}

The fidelity is obtained from the squared moduli of the two transition amplitudes, as in Eq.~\eqref{equation:Fidelity} of the main text:
\begin{align}
F = |A_g|^2 + |A_e|^2.
\end{align}
Expanding $A_{g,e} = A_{g,e}^{(0)} + A_{g,e}^{(1)}$ into a zeroth order term independent of $r_2$ and a first order term that depends on $r_2$ linearly, we obtain
\begin{align}
|A_{g,e}|^2 = |A_{g,e}^{(0)}|^2 + 2\mathrm{Re}\!\left\{A_{g,e}^{(0)} (A_{g,e}^{(1)})^*\right\} + \mathcal{O}(r_2^2).
\end{align}
Hence, the fidelity separates into a zeroth-order contribution and interference terms linear in $r_2$. The zeroth-order contribution simplifies to
\begin{align}
|A_g^{(0)}|^2 + |A_e^{(0)}|^2
= \frac{\tanh^{2N} r_1}{\cosh^2 r_1},
\end{align}
which corresponds to the fidelity obtained from a single pulse without any subsequent interactions.
The fidelity after the complete two-pulse sequence is then given by:
\begin{align}
F(\hat{\rho}_s,|N\rangle)
&= \frac{\tanh^{2N} r_1}{\cosh^2 r_1} + 2\mathrm{Re}\!\left\{A_{g}^{(0)} (A_{g}^{(1)})^* + A_{e}^{(0)} (A_{e}^{(1)})^*\right\}\\
&= \frac{\tanh^{2N} r_1}{\cosh^2 r_1}
\Bigl[1 + 2 r_2 \cos(\phi_1 - \phi_2)\,\mathcal{S}\Bigr],
\label{equation_Appendix:2pulse_scenario_final_results}
\end{align}
with
\begin{align}
\mathcal{S} &=  
\begin{aligned}[t]
\Bigl[ &\tanh^{-1} r_1 \, S_N S_{N-1} \sqrt{N(N-1)} \\
&- \tanh r_1 \, S_N S_{N+1} \sqrt{N(N+1)} \\
&+ \tanh^{-1} r_1 \, C_N C_{N-1} N \\
&- \tanh r_1 \, C_N C_{N+1} (N+1)
\Bigr]\,,
\end{aligned}
\end{align} 
which corresponds to Eq.~\eqref{equation:perturbative_two_pulse_system} in the main text. From Eq.~\eqref{equation_Appendix:2pulse_scenario_final_results} it is clear that the fidelity is always maximized if the phase difference between the pulses is $0$ or $\pi$, depending on the sign of $\mathcal{S}$.
\clearpage
\onecolumngrid
\begin{table*}[htbp!]
    \centering
        \caption{Optimal parameters for three pulses.}
    \label{tab:optimal_parameters_3p}
    \begin{tabular}{c|c|c|c|c|c|c|c|c|c}
     $N$ & $F$ & $r_1\,[\text{dB}]$   & $\phi_1$ & $t_1\,[\frac{2\pi}{\Omega}]$ & $r_2\,[\text{dB}]$ & $\phi_2$ & $t_2\,[\frac{2\pi}{\Omega}]$ & $r_3\,[\text{dB}]$   & $\phi_3$ \\ \hline 
     1 & 0.98 & 4.76 & 0 & 1.11 & 12.86 & $\pi$ & 0.19 & 12.39 & 0 \\ \hline 
    2 & 0.93 & 8.10 & 0 & 1.41 & 12.86 & $\pi$ & 0.34 & 10.96 & 0 \\ \hline 
3 & 0.84 & 9.53 & 0 & 1.49 & 13.34 & $\pi$ & 0.50 & 10.00 & 0 \\ \hline 4 & 0.74 & 9.53 & 0 & 1.49 & 13.34 & $\pi$ & 0.65 & 9.53 & 0\end{tabular}
\end{table*}
\begin{table*}[htbp!]
    \centering
        \caption{Optimal parameters for four pulses.}
    \label{tab:optimal_parameters_4p}
     \begin{tabular}{c|c|c|c|c|c|c|c|c|c|c|c|c}
     $N$ & $F$ & $r_1\,[\text{dB}]$   & $\phi_1$ & $t_1\,[\frac{2\pi}{\Omega}]$ & $r_2\,[\text{dB}]$ & $\phi_2$ & $t_2\,[\frac{2\pi}{\Omega}]$ & $r_3\,[\text{dB}]$   & $\phi_3$ & $t_3\,[\frac{2\pi}{\Omega}]$ & $r_4\,[\text{dB}]$ & $\phi_4$\\ \hline 
     1 & 0.9999 & 12.63 & 0 & 0.27 & 11.34 & $\pi$ & 1.15 & 2.84 & 0 & 0.49 & 3.47 & $\pi$ \\ \hline 
    2 & 0.9899 & 8.57 & $\pi$ & 1.20 & 3.58 & 0 & 0.27 & 11.03 & 0 & 0.25 & 12.23 & $\pi$ \\ \hline 
    3 & 0.9136 & 10.11 & $\pi$ & 1.33 & 4.05 & 0 & 0.20 & 10.86 & 0 & 0.39 & 11.41 & $\pi$ \\ \hline 
 4 & 0.8559 & 5.45 & $\pi$ & 0.65 & 7.25 & $\pi$ & 1.25 & 15.49 & 0 & 0.53 & 11.58 & $\pi$ \end{tabular}
\end{table*}
\begin{table*}[htbp!]
    \centering
        \caption{Optimal parameters for five pulses.}
    \label{tab:optimal_parameters_5p}
     \begin{tabular}{c|c|c|c|c|c|c|c|c|c|c|c|c|c|c|c}
     $N$ & $F$ & $r_1\,[\text{dB}]$   & $\phi_1$ & $t_1\,[\frac{2\pi}{\Omega}]$ & $r_2\,[\text{dB}]$ & $\phi_2$ & $t_2\,[\frac{2\pi}{\Omega}]$ & $r_3\,[\text{dB}]$   & $\phi_3$ & $t_3\,[\frac{2\pi}{\Omega}]$ & $r_4\,[\text{dB}]$ & $\phi_4$ & $t_4\,[\frac{2\pi}{\Omega}]$ & $r_5\,[\text{dB}]$ & $\phi_5$\\ \hline 
     1 & 1.0000 & 4.22 & $\pi$ & 0.76 & 3.81 & 0 & 0.84 & 1.67 & $\pi$ & 0.41 & 7.49 & $\pi$ & 0.21 & 8.93 & 0 \\ \hline 
    2 & 0.9968 & 14.20 & 0 & 0.26 & 15.10 & $\pi$ & 0.65 & 7.18 & 0 & 0.22 & 12.37 & $\pi$ & 0.22 & 8.30 & 0 \\ \hline 
    3 & 0.9626 & 16.70 & 0 & 0.43 & 9.26 & $\pi$ & 0.88 & 1.75 & $\pi$ & 1.23 & 8.86 & $\pi$ & 0.12 & 10.59 & 0 \\ \hline 
 4 & 0.8573 & 7.96 & 0 & 0.17 & 3.28 & $\pi$ & 0.39 & 7.73 & 0 & 1.26 & 15.50 & $\pi$ & 0.53 & 11.58 & 0\end{tabular}
\end{table*}
\begin{table*}[htbp!]
    \centering
        \caption{Optimal parameters for six pulses.}
    \label{tab:optimal_parameters_6p}
     \begin{tabular}{c|c|c|c|c|c|c|c|c|c|c|c|c|c|c|c|c|c|c}
     $N$ & $F$ & $r_1\,[\text{dB}]$   & $\phi_1$ & $t_1\,[\frac{2\pi}{\Omega}]$ & $r_2\,[\text{dB}]$ & $\phi_2$ & $t_2\,[\frac{2\pi}{\Omega}]$ & $r_3\,[\text{dB}]$   & $\phi_3$ & $t_3\,[\frac{2\pi}{\Omega}]$ & $r_4\,[\text{dB}]$ & $\phi_4$ & $t_4\,[\frac{2\pi}{\Omega}]$ & $r_5\,[\text{dB}]$ & $\phi_5$ & $t_5\,[\frac{2\pi}{\Omega}]$ & $r_6\,[\text{dB}]$ & $\phi_6$\\ \hline 
     1 & 1.0000 & 8.10 & 0 & 0.27 & 2.81 & $\pi$ & 0.39 & 2.29 & $\pi$ & 0.39 & 2.42 & $\pi$ & 0.61 & 3.59 & 0 & 0.86 & 1.38 & $\pi$ \\ \hline 
    2 & 0.9986 & 3.23 & 0 & 1.01 & 6.89 & $\pi$ & 0.49 & 11.02 & 0 & 0.56 & 8.19 & $\pi$ & 0.21 & 11.22 & 0 & 0.26 & 6.83 & $\pi$\\ \hline 
    3 & 0.9889 & 14.52 & $\pi$ & 0.36 & 9.25 & 0 & 1.05 & 9.40 & 0 & 0.47 & 9.87 & $\pi$ & 0.38 & 8.81 & 0 & 0.14 & 6.25 & $\pi$ \\ \hline 
 4 & 0.9549 & 10.37 & $\pi$ & 1.27 & 4.08 & $0$ & 0.23 & 10.23 & 0 & 0.62 & 15.14 & $\pi$ & 0.17 & 7.05 & 0 & 0.62 & 1.83 & $\pi$\end{tabular}
\end{table*}
\twocolumngrid

\subsection{Parameters}

Table~\ref{tab:optimal_parameters_3p} shows the 3-pulse parameters from \cite{Krstic2024}. Tables~\ref{tab:optimal_parameters_4p},
\ref{tab:optimal_parameters_5p}, and
\ref{tab:optimal_parameters_6p} show the results obtained from the gradient-based optimizations for different pulse numbers. For all optimizations, the Adam optimizer was used with hyperparameters $\alpha = 0.02, \beta_1=0.9, \beta_2= 0.999$. The Hilbert space consisted of a basis with
a maximum photon number of $d = 60$ in each mode. Among the optimization results, the optimal configuration was chosen as the highest-fidelity solution exhibiting stable convergence with respect to increasing $d$.

\subsection{Optimization Results}

\begin{figure}[htbp!]
    \centering
    \includegraphics[width=1\linewidth]{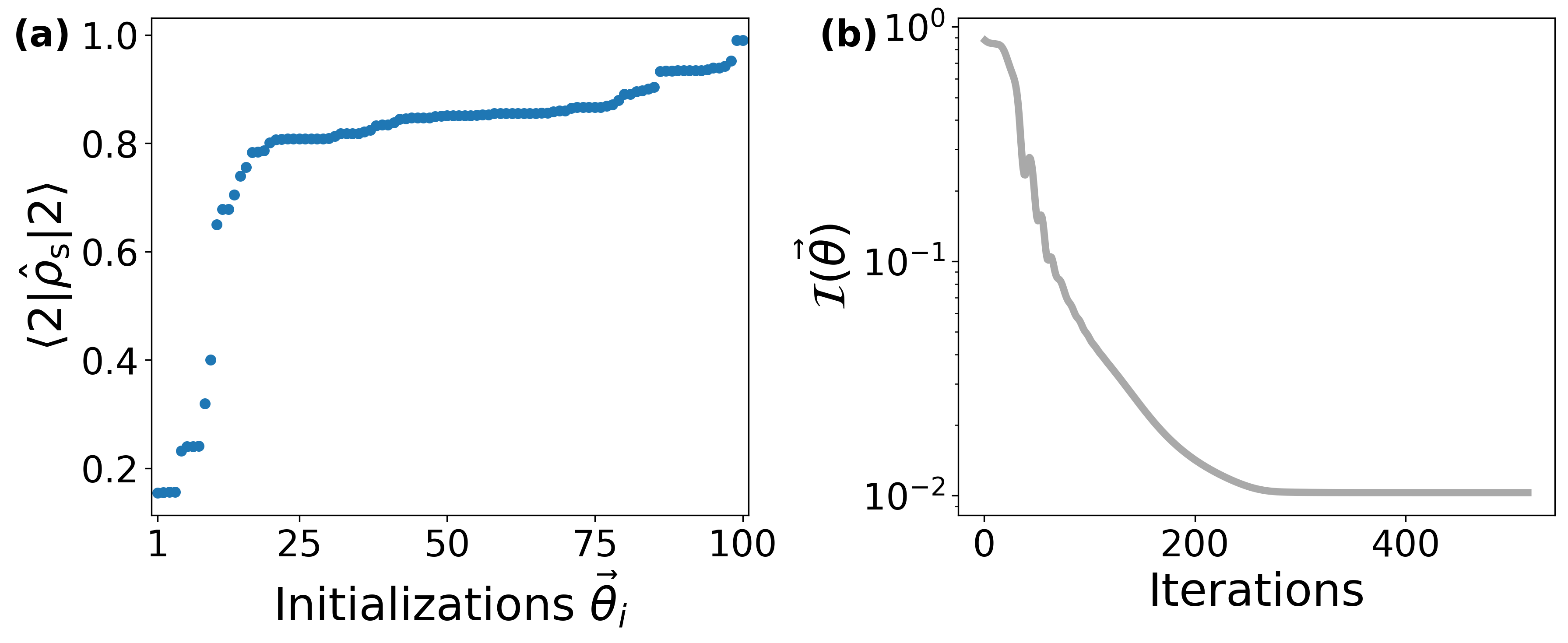}
    \caption{(a) Final fidelities achieved from 100 random initializations of the four-pulse optimization. (b) Convergence of the infidelity during the gradient-descent iterations for the run achieving the highest final fidelity.}
    \label{fig:appendix_optimization_results_N2p4}
\end{figure}

Figure~\ref{fig:appendix_optimization_results_N2p4} presents the outcomes of an exemplary gradient-descent optimization for 100 randomly initialized runs in the unitary two-photon Fock-state optimization with four pulses. The fidelities $\langle 2\lvert \hat{\rho}_\mathrm{s}\lvert 2\rangle$ resulting from the runs for different initializations $\vec{\theta}_i$ are shown in ascending order. We observe that a substantial fraction of the runs terminate at sub-optimal values, highlighting the non-convexity of the optimization landscape which renders the optimization challenging. The figure also shows the convergence curve of the cost function for the run yielding the highest-fidelity optimum. Convergence is typically reached after a few hundred iterations. Together with a maximum number of iterations, we employed an early stopping criterion as $\lVert \nabla \mathcal{I} \rVert <\epsilon $, with a typical value of $\epsilon = 10^{-3}$.

\section{Optimization Under Loss}
\label{app:opt_loss}

\subsection{Gradients}

The channel $\mathcal{E}_\text{NL}(r,\phi)( \hat{\rho}) = \hat{U}_\text{NL}(r, \phi)\hat{\rho}\, \hat{U}^\dagger_\text{NL}(r,\phi)$ describing the pulse-crystal interaction changes with the squeezing parameter $r$ as:
\begin{align}
\partial_{r}\!\left[
\hat U_{\text{NL}}\hat{\rho} \hat U_{\text{NL}}^{\dagger}
\right]
=
-i\bigl(
\mathrm{e}^{\mathrm{i}\phi}\hat a_i^{\dagger}\hat a_s^{\dagger}
+\mathrm{e}^{-\mathrm{i}\phi}\hat a_i\hat a_s
\bigr)
\hat U_{\text{NL}}\hat{\rho} \hat U_{\text{NL}}^{\dagger}
+\text{h.c.}\,.
\end{align}
The channel $\mathcal{E}_\text{2LS}(t_j)(\hat{\rho}) = \mathrm{e}^{\mathcal{L}t_j}(\hat{\rho})$ describing inter-pulse delays, during which any desired loss mechanism is described via the Lindbladian $\mathcal{L}$, changes with the time-delay parameter $t_j$ as:
\begin{align}
    \partial_{t _j}\mathrm{e}^{\mathcal{L}t_j}(\hat{\rho}) = \mathcal{L} \Bigl(\mathrm{e}^{\mathcal{L}t_j}(\hat{\rho})\Bigr)\,.
\end{align}
Thus, the derivative can be computed directly from the right-hand side of the LME:
\begin{align}
\partial_{t_j} \mathrm{e}^{\mathcal{L} t_j}(\hat{\rho})
=
&-\mathrm{i}[\hat{H}_\mathrm{JC}, \mathrm{e}^{\mathcal{L} t_j}(\hat{\rho})]\nonumber \\
 &+
\hat{L} \mathrm{e}^{\mathcal{L} t_j}(\hat{\rho}) \hat{L}^\dagger
- \tfrac{1}{2}\{ \hat{L}^\dagger \hat{L}, \mathrm{e}^{\mathcal{L} t_j}(\hat{\rho}) \}
\,,
\end{align}
where $\hat{H}_\mathrm{JC}$ is the Jaynes-Cummings Hamiltonian and $\hat{L}$ denotes the jump operator which is given as $\hat{L} = \sqrt{\gamma_\mathrm{a}}\hat{\sigma}$ in the case of atom decay and $\hat{L} = \sqrt{\gamma_\mathrm{s}}\hat{a}_\mathrm{s}$ for photon loss in the signal mode. The exponential map $\mathrm{e}^{\mathcal{L} t_j}(\hat{\rho})$ is calculated for a time-interval $t_j$ using a Trotterization which we explain in the following.
\subsection{Trotterized Decomposition of Dissipative Channels}

To simulate the dissipative dynamics efficiently, we employ a Trotterized decomposition of the Lindblad channels. 
The evolution under a Lindbladian $\mathcal{L}$ over a time $t_j$,
\begin{align}
    \mathrm{e}^{\mathcal{L} t_j}(\hat{\rho}) \,,
\end{align}
is divided into $n_{\Delta t} = t_j/\Delta t$ steps of size $\Delta t$:
\begin{align}
    \mathrm{e}^{\mathcal{L} t_j} = \left( \mathrm{e}^{\mathcal{L} \Delta t} \right)^{n_{\Delta t}}.
\end{align}

For a single time step, we separate the Lindbladian into a coherent part $\mathcal{L}_\Omega$ 
(describing the idler–2LS interaction) and a dissipative part $\mathcal{L}_\gamma$ (describing decay channels):
\begin{align}
    \mathrm{e}^{(\mathcal{L}_\Omega + \mathcal{L}_\gamma)\Delta t}(\hat{\rho}) 
    \approx \mathrm{e}^{\mathcal{L}_\Omega \Delta t} \circ \mathrm{e}^{\mathcal{L}_\gamma \Delta t}(\hat{\rho}).
\end{align}
The coherent channel can be implemented as a unitary transformation,
\begin{align}
    \mathrm{e}^{\mathcal{L}_\Omega \Delta t}(\hat{\rho}) = \hat U_\text{2LS}(\Delta t)\, \hat{\rho} \, \hat U_\text{2LS}^\dagger(\Delta t),
\end{align}
while the dissipative part of the Trotterization is modeled using the appropriate quantum channels. Specifically, atom decay is incorporated via the transformation defined in Eq.~\eqref{equation:pure_atom_decay_channel} of the main text, while photon loss in the signal mode is represented through the operator-sum decomposition given in Eq.~\eqref{equation:pure_photon_loss_channel}.

The error $\nu$ due to non-commuting contributions is of order $\Delta t^2$:
\begin{align}
    \nu = \mathrm{e}^{(\mathcal{L}_\Omega + \mathcal{L}_\gamma)\Delta t} - \mathrm{e}^{\mathcal{L}_\Omega \Delta t} \mathrm{e}^{\mathcal{L}_\gamma \Delta t} = \frac{\Delta t^2}{2}[\mathcal{L}_\Omega,\mathcal{L}_\gamma] + \mathcal{O}(\Delta t^3).
\end{align}

To improve accuracy, we use a symmetric Trotter scheme:
\begin{align}
    \mathrm{e}^{(\mathcal{L}_\Omega + \mathcal{L}_\gamma)\Delta t} \approx 
    \mathrm{e}^{\mathcal{L}_\gamma \Delta t/2} \, \mathrm{e}^{\mathcal{L}_\Omega \Delta t} \, \mathrm{e}^{\mathcal{L}_\gamma \Delta t/2} \,,
\end{align}
which reduces the leading error to $\mathcal{O}(\Delta t^3)$ while maintaining the same computational complexity. 
Half time steps involving $\mathcal{L}_\gamma$ are applied only at the start and end of the evolution; all other steps remain unchanged.

This decomposition, combined with a tailored Hilbert-space truncation, allows efficient numerical simulation and optimization of dissipative dynamics.
\subsection{Basis truncation}
The Hilbert space truncation was done by omitting states that acquire negligible occupation within the relevant timescale. The atom decay channel and Jaynes-Cummings interaction couple states sequentially, i.e., 
$|n-1,n,e\rangle \to \lvert n-1,n,g\rangle \to \lvert n-2,n,e\rangle \to \dots  \lvert 0,n,g\rangle$. Constructing the full basis would require $(d+1)^2$ states for a maximum photon number $d$. However, it suffices to include only states that can be reached within a finite number of decay steps. Retaining only states up to the order $n_\text{cut}$, the basis used for numerical computations consists of the states: 
\begin{align} 
\nonumber \Bigl\{&\lvert n,n,g\rangle, \lvert n-1,n,e\rangle,\\
\nonumber &\lvert n-1, n, g\rangle, \lvert n-2, n, e\rangle,...\\
 &\lvert n-n_\text{cut}, n, e\rangle, \lvert n-n_\text{cut}, n, g\rangle\Bigr\}\,.
\end{align}
In the case of photon loss in the signal mode, the basis consists of the states:
\begin{align} 
\nonumber \Bigl\{&\lvert n,n,g\rangle, \lvert n-1,n,e\rangle,\\
\nonumber &\lvert n, n-1, g\rangle, \lvert n-1, n-1, e\rangle,...\\
 &\lvert n, n-n_\text{cut}, g\rangle, \lvert n-1, n-n_\text{cut}, e\rangle\Bigr\}\,.
\end{align}
\subsection{Optimization Hyperparameters}
The optimization targeting the single-photon Fock state was done using a trotter step $\Delta t = 0.025\,\Omega^{-1}(0.01\,\Omega^{-1})$ under atom decay (photon loss) for decay rates of $\gamma_\mathrm{a} = 0.05\,\Omega$ and $\gamma_\mathrm{s} = 0.01\,\Omega$, respectively. 
The Hilbert space consisted of a basis with a maximum photon number of $d=60$ in each mode and a cutoff of states in the loss sector $n_\text{cut}=3$ (in both cases). The two-photon target state optimization under photon loss was done using a trotter step $\Delta t = 0.01\,\Omega^{-1}$, $n_\text{cut}=3$, $d=60$, once with $\gamma_s = 0.01\,\Omega$ and once with $\gamma_s = 0.03\,\Omega$. For all optimizations under dissipation the Adam hyperparameters were set to $\alpha = 0.05, \beta_1=0.9, \beta_2= 0.999$.
\onecolumngrid
\subsection{Parameters}
The optimal parameter configurations obtained from the loss-aware optimizations are presented in Table \ref{tab:optimal_parameters_under_dissipation_4p}. The fidelity was computed using a similar Trotterization scheme as in the optimization, i.e. $\Delta t = 0.01\,\Omega$, $n_\mathrm{cut} = 3$, and $d = 60$. To rule out numerical artefacts, we performed convergence tests with respect to $d$, $n_\mathrm{cut}$, and $\Delta t$. While convergence depends on the specific parameter constellation and decay rate, the maximum deviation of the fidelity between calculations using converged truncation-hyperparameters and those used during optimization is less than $5 \times 10^{-3}$ for the configurations shown in Table \ref{tab:optimal_parameters_under_dissipation_4p}.
\begin{table}[htbp!]
    \centering
        \caption{Optimal parameters under dissipative conditions for four pulses.}
    \label{tab:optimal_parameters_under_dissipation_4p}
     \begin{tabular}{c|c|c|c|c|c|c|c|c|c|c|c|c|c|c}
     $N$ & $F$ & decay mechanism & decay rate & $r_1\,[\text{dB}]$   & $\phi_1$ & $t_1\,[\frac{2\pi}{\Omega}]$ & $r_2\,[\text{dB}]$ & $\phi_2$ & $t_2\,[\frac{2\pi}{\Omega}]$ & $r_3\,[\text{dB}]$   & $\phi_3$ & $t_3\,[\frac{2\pi}{\Omega}]$ & $r_4\,[\text{dB}]$ & $\phi_4$\\
     \hline 
     1 & 0.965 & atom decay & $\gamma_\mathrm{a} = 0.05\,\Omega$ & 18.52 & 0 & 0.12 & 15.62 & $\pi$ & 0.61 & 3.75 & $\pi$ & 0.20 & 4.90 & 0\\\hline 
      1 & 0.815 & photon loss & $\gamma_\mathrm{s}=0.03\,\Omega$ & 5.93 & $\pi$ & 0.67 & 2.55 & $0$ & 0.44 & 7.82 & 0 & 0.24 & 8.35 & $\pi$\\
      \hline 
      2 & 0.784 & photon loss & $\gamma_\mathrm{s} = 0.01\,\Omega$ & 8.43 & $\pi$ & 1.21 & 3.84 & 0 & 0.26 & 8.91 & 0 & 0.32 & 9.96 & $\pi$ \\
      \hline 
      2 & 0.531 & photon loss & $\gamma_\mathrm{s} = 0.03\,\Omega$ & 6.87 & 0 & 1.12 & 3.56 & $\pi$ & 0.26 & 8.24 & $\pi$ & 0.41 & 8.06 & 0 
    \end{tabular}
\end{table}
\end{document}